\documentclass[aps,prb,amsmath,amssymb,twocolumn,superscriptaddress,showpacs,floatfix]{revtex4-1}

\usepackage{bm}
\usepackage{epsfig}
\usepackage[usenames]{color}
\usepackage{graphicx}
\usepackage{hyperref}
\usepackage{esint}

\usepackage{eurosym}

\newcommand{\beq}{\begin{equation}}
\newcommand{\eeq}{\end{equation}}
\newcommand{\beqa}{\begin{eqnarray}}
\newcommand{\eeqa}{\end{eqnarray}}

\newcommand{\Em}{\mathcal E}

\DeclareMathOperator{\Imm}{Im}

\DeclareMathOperator{\Rem}{Re}

\def\XXint#1#2#3{{\setbox0=\hbox{$#1{#2#3}{\int}$ }
\vcenter{\hbox{$#2#3$ }}\kern-.5\wd0}}

\newcommand*{\pvint}
{
  \mathop{\,\,\vphantom{\intop}\!\!\!
  \mathpalette\pvintop\relax}\nolimits}
\newcommand*{\pvintop}
[2]{
  \ooalign{$#1\intop$\cr\hidewidth$#1-$\hidewidth}}
\newcommand*{\PVI}{\pvint}

\begin{document}

\date{\today }

\title{Low-Bias-Anomaly and Tunnel Fluctuoscopy}

\author{A.~Glatz}
\affiliation{Materials Science Division, Argonne National Laboratory, Argonne, Illinois 60439, USA}
\affiliation{Department of Physics, Northern Illinois University, DeKalb, Illinois 60115, USA}

\author{A.\,A.\,Varlamov}
\affiliation{CNR-SPIN, Viale del Politecnico 1, I-00133 Rome, Italy}
\affiliation{Materials Science Division, Argonne National Laboratory, Argonne, Illinois 60439, USA}

\author{V.\,M.\,Vinokur}
\affiliation{Materials Science Division, Argonne National Laboratory, Argonne, Illinois 60439, USA}

\begin{abstract}
Electron tunneling spectroscopy pioneered by Esaki\cite{esaki} and Giaever\cite{giaver,GM61} offered a powerful tool for studying electronic spectra and density of
states (DOS) in superconductors.
This led to important discoveries that revealed, in particular,
the pseudogap in the tunneling spectrum of superconductors above their
critical temperatures\cite{ARW70,VD83,CCRV90,Benjamen10}. 
However, the phenomenological approach of Ref.~\cite{GM61} 
does not resolve the fine structure of low-bias behavior
carrying significant information about electron scattering,
interactions, and decoherence effects. 
Here we construct a complete microscopic theory of electron tunneling 
into a superconductor in the fluctuation regime. 
We reveal a non-trivial low-energy anomaly in tunneling conductivity due to 
Andreev-like reflection of injected electrons from superconducting fluctuations.
Our findings enable real-time observation of fluctuating Cooper pairs  dynamics by time-resolved scanning tunneling microscopy measurements and open new  horizons for quantitative analysis of the fluctuation electronic spectra of superconductors.
\end{abstract}

\maketitle

There have been rapid developments in scanning tunneling microscopy
(STM) or scanning tunneling spectroscopy (STS) studies of superconductivity
triggered by investigations of the pseudogap state and vortex state in
high-temperature cuprates\cite{micklitz+prb09}, observations of the
pseudogap in 2D disordered films of conventional superconductors~\cite%
{Benjamen10}, investigations of the superconductor-insulator transition~\cite%
{Sacepe}, measurements of the tunnel conductivity close to the
superconducting transition in intrinsic Josephson junctions
(see Ref. \cite{K11}), and many others. All this called for a quantitative
theory capable to adequately describe high resolution STM/STS data
uncovering subtle features of the tunneling spectra. Of special importance
is the ability of analyzing data in the fluctuation regime as it is the
domain that is key to reveal the microscopic mechanisms of high temperature
superconductivity and the superconductor-insulator transition.

However, the restrictions of the phenomenological GM approach disguise
the fine structure of the electronic spectrum. To see how this is
happening, let us inspect the classical GM expression\cite{GM61} for the tunnel current
\begin{eqnarray}
I_{\mathrm{qp}}\left( V\right) &=&-\frac{\hbar }{eR_{N}\nu _{L}\left(
0\right) \nu _{R}\left( 0\right) }\int_{-\infty }^{\infty }\left[
n_{F}\left( E+eV\right) -n_{F}\left( E\right) \right]  \notag \\
&&\cdot \nu _{L}\left( E+eV\right) \nu _{R}\left( E\right) dE,\,
\label{giaver}
\end{eqnarray}%
%

\begin{figure}[tbh]
\centering
\includegraphics[width=0.45\columnwidth]{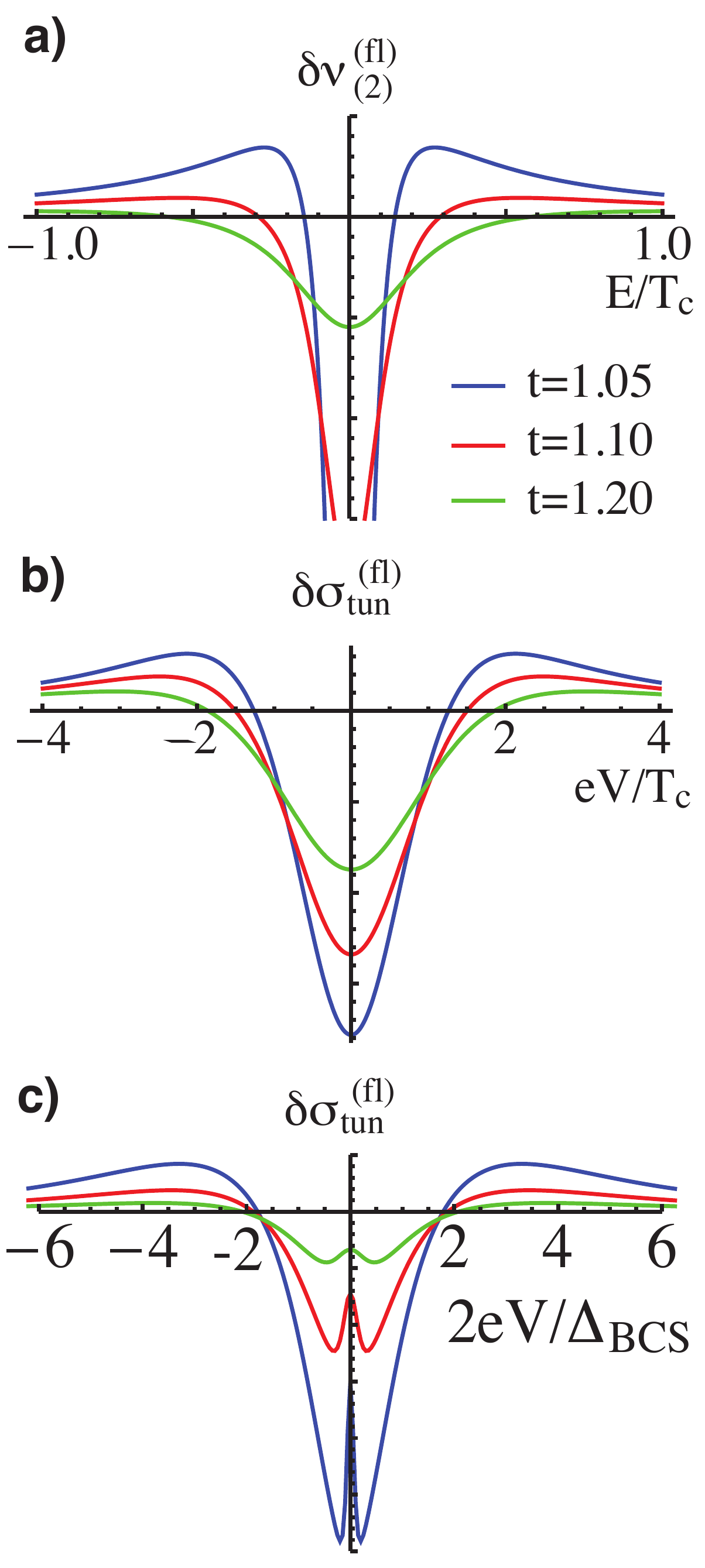}
\caption{\textbf{a)} Theoretical curves of the fluctuation correction to the
single particle DOS, $\protect\delta \protect\nu _{(2)}^{\mathrm{(fl)}}$,
versus energy, $E$, for 2D superconductors above the critical temperature
for temperatures close to the critical one ($t=T/T_{c0}=1.05,1.1,1.2).$
showing a pronounced divergence at zero energy. \textbf{b)} The resulting
pseudogap in the tunneling conductivity obtained by applying the GM Eq. (1)
to the fluctuation correction to the $\protect\delta \protect\nu _{(2)}^{%
\mathrm{(fl)}}$. \textbf{c)} The low-voltage anomaly of the tunneling
conductivity related to Andreev-like reflection of injected electron from the
fluctuating superconductive domain, which is beyond the possibilities of the
GM approach.}
\label{fig.ados}
\end{figure}
\begin{figure*}[tbh]
\begin{center}
\includegraphics[width=\textwidth]{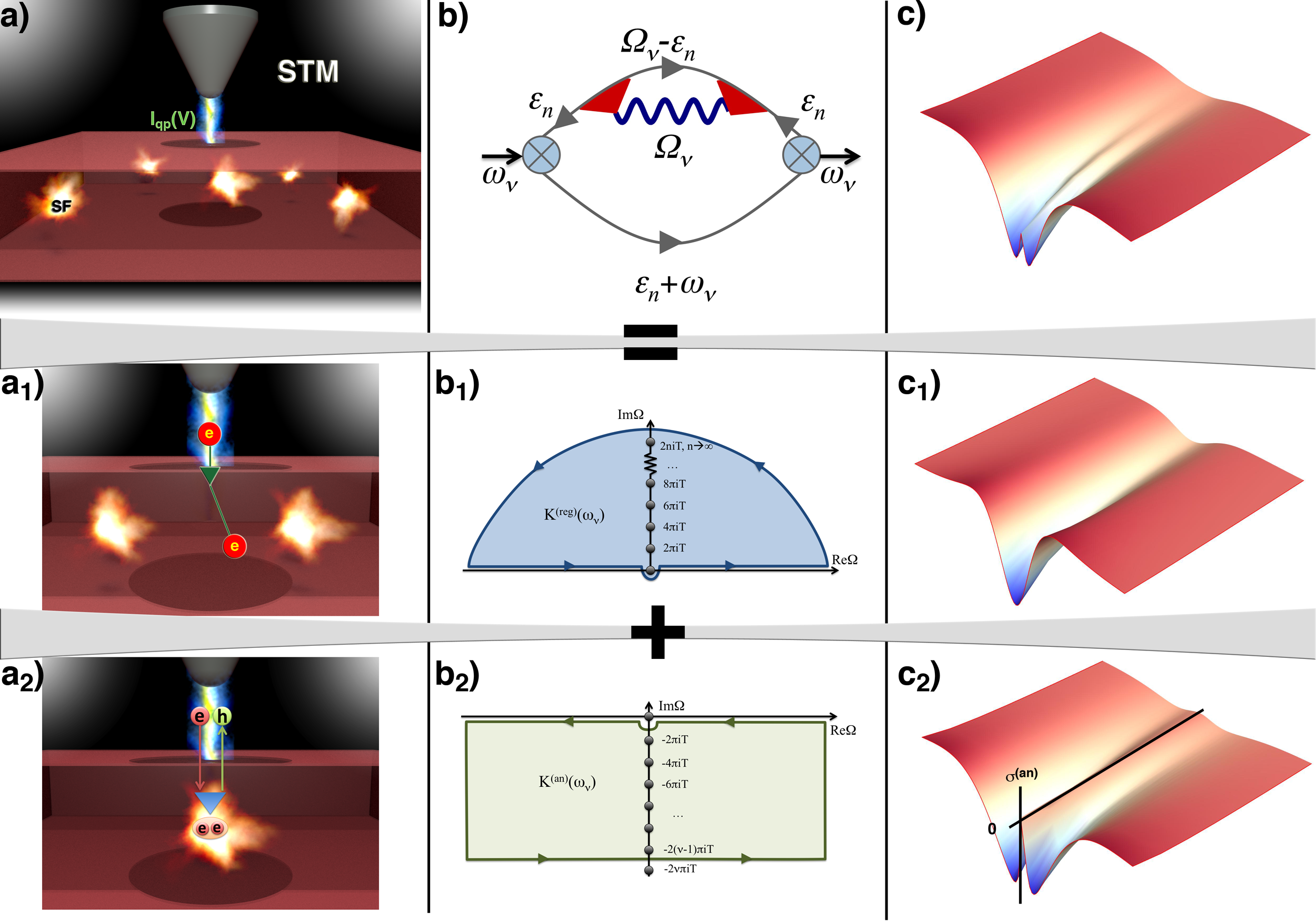}
\end{center}
\caption{\textbf{a}). Schematic STM setup of a N-I-(N+SF) tunnel experiment,
\textbf{a$_{1}$}). An injected electron pair (2e) thermalizes in the
electrode, which reduces the density of states due to superconducting
fluctuations, \textbf{\ $a_{2}$}). Andreev-like reflection of injected electrons
at a region of superconducting fluctuations (SF); \textbf{b}). The
(Matsubara) diagram describing the fluctuation contribution to tunneling
current, \textbf{b$_{1}$})+\textbf{b$_{2}$}) Two contours in the plane of
complex voltage describing both both corresponding tunneling processes shown
in a$_{1}$) and a$_{2}$); \textbf{c}). Surface plot of the total tunnel
conductivity depending on voltage and temperature. The corresponding
theoretical expression is valid throughout the whole phase diagram of
temperature and magnetic field with a wide pseudogap structure and narrow
low-bias anomaly (LBA), \textbf{c$_{1}$}). Pseudogap anomaly related to the
renormalization of the one-electron density of states due to superconducting
fluctuations in the electrode. It directly corresponds to the process
pictured in a$_{1}$) and contour b$_{1}$), \textbf{c$_{2}$}). LBA
contribution of the tunnel conductivity due to process a$_{2}$), resulting
from contour b$_{2}$).}
\label{fig.all}
\end{figure*}
(here $R_{N}$ is the tunnel junction resistance, $n_{F}\left( E\right) $ is
the Fermi distribution function, and $\nu _{L,R}$ is the energy dependent
density of states of the left (right) electrode, respectively) and apply it
to the calculation of the tunnel current of a N-I-(N+SF) junction at
temperatures above $T_{\mathrm{c}}$. Using the explicit expression for the
fluctuation correction to the electronic DOS in a disordered superconducting
film \cite{ARW70} (shown in the Fig.\thinspace\ref{fig.ados}a), one sees
with surprise that the sharp singularity in the DOS at low energies gets smoothened
out to a much wider pseudogap structure in the differential conductivity. 
In particular, the latter has the width $\delta (eV_{\mathrm{pg}})\sim \Delta _{%
\mathrm{BCS}}$ [instead of $\delta E_{0}\sim k_{B}\left( T-T_{c0}\right) $]
and a small amplitude 
[$\ln \left[ T_{c0}/(T-T_{c0})\right]$ instead of $T_{c0}^{2}/(T-T_{c0})^{2}$] in the DOS \cite{VD83} (see Fig.\thinspace \ref%
{fig.ados}\textbf{b}). The reason for these dissimilarities is that the
sign-change of the DOS fluctuation correction (see Fig.\thinspace \ref%
{fig.ados}\textbf{a}), almost averages out the whole effect of fluctuations on the tunnel
current when integrated over energy. 
As a result, quantum coherent effects like Andreev reflections of injected electrons at domains of
superconducting fluctuations in the biased electrode (see
Fig.\thinspace \ref{fig.ados}\textbf{c}) cannot be described by the GM phenomenology.

To construct a general approach to calculate the true tunnel conductivity taking into account the fine structure of the density of states,
we employ the Matsubara Green functions technique. 
The complete fluctuation contribution to the tunneling current in a typical STM/STS experiment
sketched in Fig.\ref{fig.all}\textbf{a}
is represented graphically by the
Matsubara diagram, shown in Fig.\ref{fig.all}\textbf{b}. 
This diagram describes both, regular and anomalous fluctuation tunneling
processes depicted in Fig.\ref{fig.all}\textbf{a$_1$} and Fig.\ref{fig.all}%
\textbf{a$_2$}. 
The former one, related to the depletion of the electron DOS close
to the Fermi level, has already been discussed above in the framework of the
phenomenological theory~\cite{GM61} and, as we know, results in the appearance
of the pseudogap-like feature in $\delta \sigma _{\mathrm{tun}}^{\left(
\mathrm{fl}\right) }\left( V\right)$. 
The latter process consists of Andreev-like
reflections of an injected, still energetically unrelaxed, electron from the
fluctuation superconducting domain in the biased electrode, shown in Fig.\ref%
{fig.all}\textbf{a$_2$}. In order to participate in fluctuation Cooper
pairing, the injected electron ``extracts'' an electron-hole pair from
vacuum with momentum opposite to its own, forms a Cooper pair with the
electron, while the remaining hole returns along its previous trajectory
(see Fig.\,\ref{fig.all}\textbf{a$_2$}). This quantum coherent contribution
is missed by the phenomenological method, but is captured by the microscopic diagrammatic
approach.
This anomalous tunneling process gives rise to an additional
current, which, like the regular one, is proportional to the first power of the
Ginzburg number $\mathrm{Gi}$ (which characterizes the strength of
fluctuations), but is cubic in voltage $V$ near zero bias and becomes
relevant only close enough to the superconducting transition. 
As a result, a peculiar \textit{low-bias anomaly} (LBA) appears near the superconducting
transition line $H_{c2}(T)$.  As the  external
parameter values move away from the transition line the amplitude of the LBA rapidly decays.
The important feature of this novel Andreev process is that it appears
in lowest (first) order approximation with respect to the tunneling
barrier transparency -- the same order as the usual tunneling current exhibiting the pseudogap. 
This effect is stronger than the standard Andreev conductance of a N-I-S junction
which is proportional to the square of the transparency\cite{BTK82,HK95}.
The reason is that the fluctuation-induced domain of superconducting phase
in the biased electrode is not separated from the surrounding normal phase by
any barrier and thus the process of Andreev-like reflection does not involve an
additional tunneling process.

\begin{figure*}[tbh]
\begin{center}
\includegraphics[ width=\textwidth]{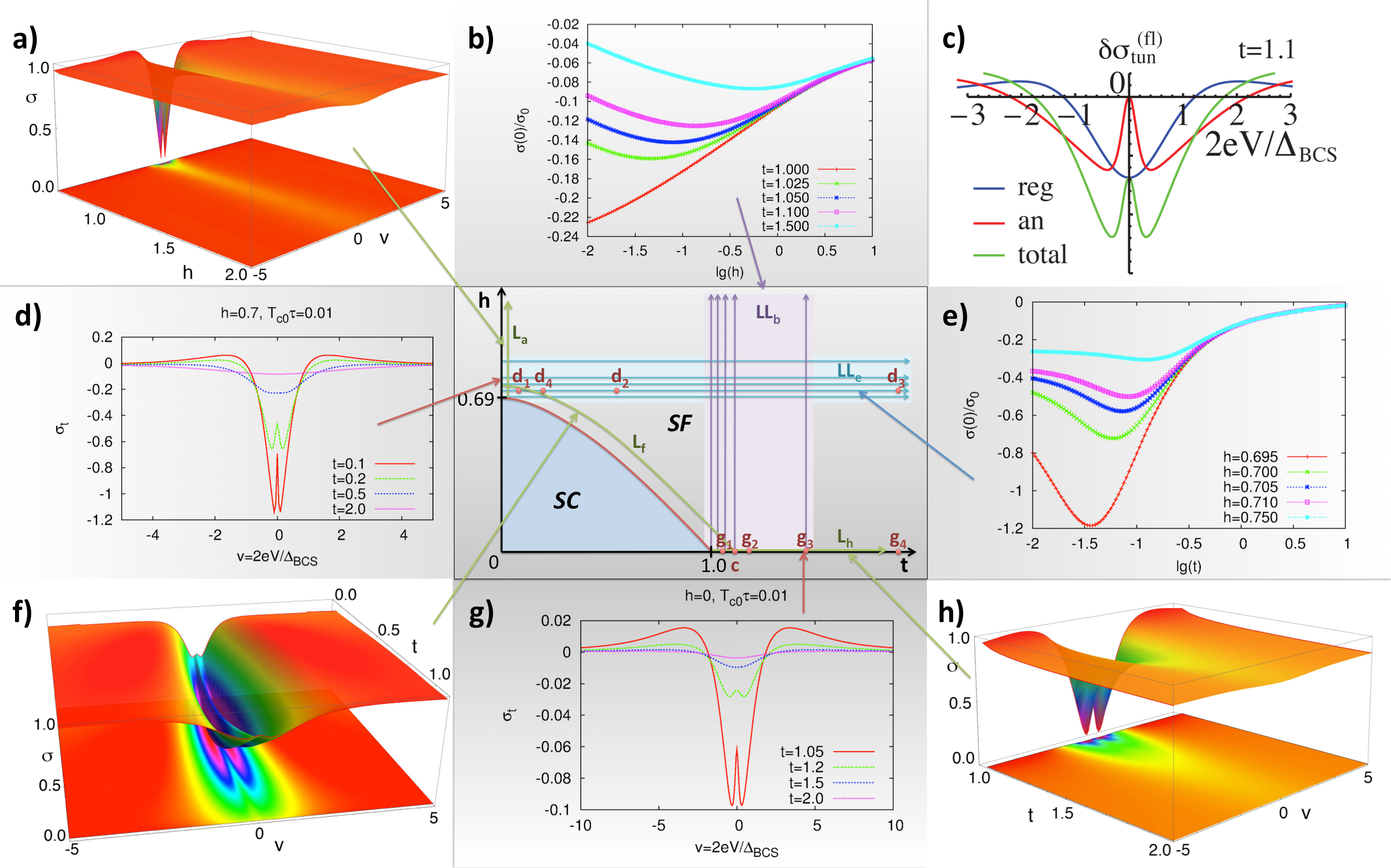}
\end{center}
\caption{Various plots of the tunneling conductance for different cuts and
point in the $t-h$ plane. The cut lines and points are indicated in the $t-H$
phase diagram in the central panel. Points are labeled by the panel letter, lines by ``L'' and panel letter subscript. \textbf{a)} Low temperature ($t=0.05$)
dependence of the conductivity as surface plot depending on voltage, $v$,
and magnetic field, $h>h_{c2}(0)=0.69$ [cut line L$_{a}$]. \textbf{b)}
Zero-bias conductivity at fixed temperatures as function of $\ln (h)$ [cut
lines LL$_{b}$]. \textbf{c)} $t=1.1$ plot of the components (pseudo gap, ``reg'', and LBA, ``an'') of the tunnel conductivity [point c]. \textbf{d)} Tunnel conductance for $h=0.7$ at different
temperatures depending on $v$ [points d$_{1}$-d$_{4}$]. \textbf{e)}
Zero-bias conductivity at fixed magnetic field as function of $\ln (t)$ [cut
lines LL$_{e}$]. \textbf{f)} Conductivity as surface plot depending on
voltage and closely following the superconducting transition line in the $%
t-h $ plane [cut line L$_{f}$].\textbf{g)} Tunnel conductance for $h=0$ at
different temperatures depending on $v$ [points g$_{1}$-g$_{4}$]. \textbf{h)} Zero field ($h=0$) dependence of
the conductivity as surface plot depending on voltage, $v$, and temperature,
$t>t_{c}=1$ [cut line L$_{h}$] (the same parameters as used for column c) of
Fig.\protect\ref{fig.all}).  }
\label{fig.plots}
\end{figure*}

Remarkably, both complimentary physical processes shown in
panels \textbf{a$_1$} and \textbf{a$_2$} of Fig.\ref{fig.all} are
straightforwardly expressed in terms of a graphic mathematical language: the
calculation of the diagram of Fig.\ref{fig.all}\textbf{b} is reduced to the
evaluation of the integrals of the electron Green functions in the linked
electrodes along two contours in the complex frequency plane shown in
panels \textbf{b$_1$} and \textbf{b$_2$} of Fig.\ref{fig.all}, respectively. 
The upper contour corresponds to the conventional Giaever-Megerle (GM) tunneling,
while the lower one describes the contribution due
to Andreev-like reflection from superconducting fluctuations. 
Accordingly, the fluctuation part of the tunneling conductance shown in Fig.\ref%
{fig.all}\textbf{c} exhibits both, the pseudogap anomaly due to
fluctuation depletion of the one-electron DOS (Fig.\ref{fig.all}\textbf{c$%
_1 $}) coming from the integration over the contour of Fig.\ref{fig.all}%
\textbf{b$_1$}, and Andreev-like reflection induced LBA ( Fig.\ref{fig.all}%
\textbf{c$_2$}), arising from the integration over the
contour of the panel \textbf{b$_2$}. Important to remark is that the latter contribution is zero at zero bias voltage [see Fig.\ref{fig.plots}\textbf{c}].

In the framework of the diagrammatic Matsubara formalism the tunneling current is presented as (see
Methods):
\begin{equation}
I_{\mathrm{qp}}\left( V\right) =-e\Imm K^{R}(\omega _{\nu }\rightarrow -ieV),
\label{Iqp}
\end{equation}%
where
\begin{equation}
K(\omega _{\nu })\!=4T\sum_{\varepsilon _{n}}\sum_{\mathbf{pk}}|T_{\mathbf{pq%
}}|^{2}G_{L}\left( \mathbf{p},\varepsilon _{n}+\omega _{\nu }\right)
G_{R}\left( \mathbf{k},\varepsilon _{n}\right) .  \label{K}
\end{equation}%
Here $G_{L}$ and $G_{R}$ are the exact Matsubara Green functions of the left
and right electrodes respectively, the summations are performed over all
fermionic frequencies $\varepsilon _{n}=2\pi T(n+1/2)$ and the electron
states $\mathbf{p}$ and $\mathbf{k}$\textbf{\ } in the corresponding
electrodes.\ The external bosonic frequency $\omega _{\nu }$ accounts for
the potential difference between the electrodes and the factor $4$ is due to
the summation over the spin degrees of freedom. The superscript
\textquotedblleft R\textquotedblright\ in Eq. (2) means that the correlator $%
K$ is continued to the plane of complex voltages in such a way that it
remains an analytic function through the complete upper complex half-plane. 

The calculation of the sums in Eq.\thinspace (\ref{K}) is presented in the
appendix.
It turns out that the discussed LBA in the I-V characteristics appears only in the case where the
energy (or phase) relaxation time $\tau _{\phi }$ of an electron injected into
the explored electrode is long enough: $T_{c0}\tau _{\phi }\gg \hbar /k_{B}.$
The shape of the LBA close to the critical temperature [$\hbar \tau _{\phi
}^{-1}\lesssim k_{B}\left( T-T_{c0}\right) \ll k_{B}T_{c0}$] for low
voltages $eV\lesssim k_{B}\left( T-T_{c0}\right)$, can be found
analytically:
\begin{equation}
\sigma _{\mathrm{tun}}^{\left( \mathrm{fl}\right) }=-\frac{7\zeta \left(
3\right) e^{2}S}{2\pi ^{4}\hbar \sigma _{n}R_{N}}\left[ \ln \frac{T_{c0}}{%
T-T_{c0}}+\frac{3\tau _{\phi }}{8\pi \hbar k_{B}}\frac{\left( eV\right) ^{2}}{%
\left( T-T_{c0}\right) }\right] ,  \label{final}
\end{equation}%
with $\sigma _{n}$ as the electrode normal conductivity and $S$ as the
junction surface area. When $k_{B}\left( T-T_{c0}\right)$ decreases to
the value $\hbar \tau _{\phi }^{-1}$ the growth of the LBA ceases. One can
show that close to the transition temperature $T_{c0}$ the dip in the tunnel
conductivity develops on the scale $eV_{\mathrm{LBA}}^{\mathrm{TF}}\sim
\Delta _{\mathrm{BCS}}^{1/2}\sqrt{\hbar \tau _{\phi }^{-1}\left(
T-T_{c0}\right) /T_{c0} }\ll \Delta _{\mathrm{BCS}}$. At
zero temperature, close to the second critical field $H_{c2}\left( 0\right) ,$
the fluctuations acquire quantum nature and the corresponding voltage scale is $eV_{%
\mathrm{LBA}}^{\mathrm{QF}}\sim \Delta _{\mathrm{BCS}}^{1/2}\sqrt{\hbar \tau
_{\phi }^{-1}\left[ H-H_{c2}\left( 0\right) \right] /
H_{c2}\left( 0\right)  }\ll \Delta _{\mathrm{BCS}}$. From the obtained Eq. (\ref{final}) one
sees, that the intensity of the LBA is directly proportional to the energy
relaxation length $\ell _{\phi }=v_{F}\tau _{\phi },$ which is in a complete
agreement with the physical picture of this non-trivial
quantum coherence effect presented above: anomalous Cooper pairings take place only in
a stripe of volume $S\cdot \ell _{\phi }$ in the contact area,
where the injected electrons still remain non-thermalized and differ from
the local ones.

Fig.\thinspace \ref{fig.plots} shows the plots of fluctuation contributions
to the tunneling conductivity for different parts of the
temperature-magnetic field phase diagram of the superconducting film. The
central panel -- the $h$-$t$ phase diagram -- depicts the parameter
combinations or ranges for the 2D graphs or 3D surface plots arranged around
it in panels a) - h). In accordance with the above theoretical speculations
the strength of the singularity in the low-voltage behavior of the tunneling
conductance smears out when moving away from the transition line (panels a-d
and g, h). We point out that the LBA is most pronounced roughly halfway
between the 'endpoints' of the transition line (see panel f). Overall the
panels clearly show that the LBA is a pronounced important effect near the
transition and is even noticeable at twice the transition temperature.

In conclusion, the LBA provides an irreplaceable tool for
determining microscopic material parameters including the energy relaxation time $\tau _{\phi
} $, the critical temperature $T_{c0}$, and the critical magnetic field $%
H_{c2}(0)$ by measuring the tunneling conductance and fitting the experimental data with
the complete expression for the tunneling conductance. Remarkably, all the information about these parameters is encoded in merely the distance
between the LBA dips and the height
of the central peak in the conductivity curve. 
This introduces a new technique, a \textit{tunnel-fluctuoscope}, in analogy to the recently developed conductivity fluctuoscopy\cite{GVV11,glatz+epl11}.
The latter has already proven to be the first quantitative and precise method to determined material parameters of superconducting films in many instances.
An observation of the described LBA in a d.c. experiment is a
fingerprint of the fact that at the point below the STM tip, FCPs appear during the time of the experiment. 
Recent tunnel-current measurements of N-I-S junctions indeed indicate the presence of the LBA\cite{Krasnov}.
Since the characteristic
FCP lifetime is $\hbar /k_{B}\left( T-T_{c}\right)$, a time-resolved STM measurement
utilizing an a.c. current with frequency on the scale of 1-10 GHz promises to make it
possible, in principle, to \textquotedblleft visualize\textquotedblright\ FCP directly in
real time.

\section{Acknowledgments}

We acknowledge useful
discussions with B.~Altshuler, A.~Goldman, A.~Kamenev, V.E.~Kravtsov, V.~Krasnov, and
M.~Norman. The work was supported
by the U.S. Department of Energy Office of Science under the Contract No.
DE-AC02-06CH11357. A.A.V. acknowledges support of the FP7-IRSES program,
grant N 236947 ``SIMTECH'' .

\appendix

\section{Calculation methods}

We study low-transparency junctions in the regime of weak fluctuations and
find the tunneling current $I\left( V\right) $ between a normal metal
electrode and a disordered two-dimensional superconducting film placed in a
perpendicular magnetic field throughout the whole phase diagram above the $%
H_{c2}(T)$ line. This system can be described by the tunnel Hamiltonian with
interaction term%
\begin{equation}
\widehat{\mathcal{H}}_{T}=\sum_{\mathbf{p,k,\sigma }}\left( T_{\mathbf{pk}}%
\widehat{a}_{\mathbf{p\sigma }}^{+}\widehat{b}_{\mathbf{k\sigma }}+T_{%
\mathbf{pk}}^{\ast }\widehat{b}_{\mathbf{k}}^{+}\widehat{a}_{\mathbf{p}%
}\right)
\end{equation}%
and the tunnel current can be identified with the time derivative of the particle number
operator in one of the electrodes%
\begin{equation}
\widehat{\mathcal{N}}_{L}=\sum_{\mathbf{k,\sigma }}\widehat{a}_{\mathbf{%
k\sigma }}^{+}\widehat{a}_{\mathbf{k\sigma }}
\end{equation}%
averaged over the statistical ensemble:%
\begin{equation}
I_{\mathrm{qp}}\left( V,T\right) =e\left\langle \frac{d\widehat{\mathcal{N}}%
_{L}}{dt}\right\rangle =-\frac{ie}{\hbar }\left\langle \left[ \widehat{%
\mathcal{N}}_{L},\widehat{\mathcal{H}}_{T}\right] \right\rangle .
\end{equation}%
The procedure of such ensemble averages with the density matrix was performed
in Ref.~\cite{R97}. The tunnel current is then determined by the diagram
presented in Fig.\ref{fig.all}\textbf{b} appearing in first orders of barrier transparency and strength of fluctuations $\mathrm{Gi}$.
Solid lines correspond to the single-electron Green's functions
in the respective electrodes, the wavy line represents the fluctuation
propagator, crossed circles stand for the matrix elements of the tunneling
Hamiltonian, and the solid triangles are the vertices
accounting for impurity averaging. The quasi-particle current flowing
through a tunnel junction is expressed via the correlator $K\left( \omega
_{\nu }\right) $ of the electron Green's functions of both
electrodes. Being calculated as a series of imaginary Matsubara
frequencies\cite{energy} $i\omega _{\nu }=2\pi iT\nu ,$ $\nu =0,1,2,..,$ the
obtained expression has to be analytically continued into the whole upper
half-plane of complex frequencies $\omega $: $i\omega _{\nu }\rightarrow
\omega $. Finally, the real positive values of the latter are identified
with the voltage at junction $\omega=eV$.

As it is shown in the appendix, after summations over momenta and
fermionic frequencies, the correlator $K^{\left( \mathrm{fl}\right) }\left(
\omega _{\nu }\right) $, corresponding to the diagram from Fig.\ref{fig.all}%
\textbf{b} becomes
\begin{align}
& K^{\left( \mathrm{fl}\right) }\left( \omega _{\nu }\right) =K^{\left(
\mathrm{reg}\right) }\left( \omega _{\nu }\right) +K^{\left( \mathrm{an}%
\right) }\left( \omega _{\nu }\right)   \label{K+K2} \\
& =\frac{8T_{c0}Sh}{\pi ^{3}\sigma _{n}R_{N}}\sum_{m=0}^{M}\left[
\sum_{k=0}^{\infty }+\sum_{k=-\nu }^{-1}\right] \frac{\left[ \mathcal{E}%
_{m}^{\prime }\left( k+2\nu \right) -\mathcal{E}_{m}^{\prime }\left(
k\right) \right] }{\mathcal{E}_{m}\left( |k|\right) }  \notag
\end{align}
The function
\begin{equation}
\mathcal{E}_{m}\left( x\right) =\ln t+\psi \left[ \frac{1+x}{2}+\frac{4h}{%
\pi ^{2}t}\left( m+\frac{1}{2}\right) \right] -\psi \left( \frac{1}{2}%
\right) 
\end{equation}%
represents the denominator of the fluctuation propagator describing the
fluctuation pairing of electrons in the normal phase of a superconductor
over a wide range of temperatures and fields~\cite{LV05}. Here $t=T/T_{c0}$
and $h=\pi ^{2}/(8\gamma _{E})H/H_{c2}(0)$ are dimensionless temperature and
magnetic field normalized by the critical temperature and the value of second
critical field respectively, $\gamma _{E}=1.78$ is the exponential Euler
constant. One can see that close to $T_{c0}$ and for weak enough magnetic
fields, $\mathcal{E}_{m}\left( -i\omega \right) $ is nothing else than the
fundamental solution of the time-dependent Ginzburg-Landau equation \cite%
{LV05}.

The summation over the bosonic Matsubara frequencies ``flowing'' through the
fluctuation propagator done by an additional analytical continuation in upper
half-plane results in the general expression for correlation function $%
K^{R}(\omega _{\nu }\rightarrow -ieV)$

\begin{equation}
K^{\left( \mathrm{reg}\right) R}\left( \omega _{\nu }\rightarrow -ieV\right)
=\frac{2T_{c0}Sh}{\pi ^{3}\sigma _{n}R_{N}}\sum_{m=0}\frac{\left[ \mathcal{E}%
_{m}^{\prime }\left( -\frac{ieV}{\pi T}\right) -\mathcal{E}_{m}^{\prime
}\left( 0\right) \right] }{\mathcal{E}_{m}\left( 0\right) }  \notag
\end{equation}%
\begin{eqnarray}
&&K^{\left( \mathrm{an}\right) R}\left( \omega _{\nu }\rightarrow -ieV
\right) =-\frac{1}{2}K^{\left( \mathrm{reg}\right) R}\left( -ieV\right)
-i\frac{T_{c0}Sh}{\pi ^{3}\sigma _{n}R_{N}}
\notag \\
&&\cdot \sinh \left( \frac{eV}{2T}\right) \sum_{m=0}\PVI\limits_{-\infty }^{\infty }\frac{\left[ \mathcal{E}%
_{m}^{\prime }\left( iz\!-\!\frac{ieV}{\pi T}\right) \!-\!\mathcal{E}%
_{m}^{\prime }\left( \!-iz\right) \right] dz\!}{\sinh \left( \pi z\right) \sinh \pi \left( z-\frac{eV}{2\pi T}%
\right) \mathcal{E}_{m}\left(
iz\right) },\notag
\end{eqnarray}%
which allows to obtain the fluctuation contribution to the tunnel current
for arbitrary temperatures, magnetic fields and voltages. The corresponding results for the conductivity are presented in Figs.~\ref{fig.all}\&\ref{fig.plots}.

\section{Deficiency of Phenomenological Model}

\emph{The effect of SFs on the DOS and corresponding pseudogap in tunnel
conductivity}. According to the microscopic BCS theory~\cite{BCS}, the
superconducting state is characterized by a gap in the normal excitation
spectrum, centered around the Fermi level, $E_{F}$, which vanishes along the
transition line $H_{c2}(T)$. However, it was predicted, as early as in 1970~%
\cite{ARW70}, that even in the normal state of a superconductor, thermal
fluctuations result in a noticeable suppression of the density of states (DOS) in
a narrow energy range around the Fermi level (see Fig.~\ref{fig.ados}a).

More specifically, in the case of a disordered thin film \cite{ARW70} the
fluctuation correction to the DOS takes form:
\begin{equation}
\frac{\delta \nu _{\left( 2\right) }^{\left( \mathrm{fl}\right) }\left(
E,T\right) }{\nu _{n}}\!\!=\!\frac{4.6\mathrm{Gi}_{\left( 2\right)
}k_{B}^{2}T^{2}}{\left( E\!-\!\frac{1}{2}\tau _{GL}^{-1}\right) ^{2}}\left[
\!\frac{E\!-\!\frac{1}{2}\tau _{GL}^{-1}}{E\!+\!\frac{1}{2}\tau _{GL}^{-1}}%
-\ln \frac{E\!+\!\frac{1}{2}\tau _{GL}^{-1}}{\tau _{GL}^{-1}}\!\!\!\right] ,
\label{deltanu2}
\end{equation}%
where $\nu _{n}$ is the electron density of the states per one spin of a
normal metal at the Fermi level, $\mathrm{Gi}_{\left( 2\right) }=1.3\hbar
^{2}/(p_{F}^{2}ld)$ is the Ginzburg-Levanyuk number characterizing the
strength of fluctuations in the film, $\tau _{GL}=\pi \hbar
/8k_{B}(T-T_{c0}) $ is so-called Ginzburg-Landau time, characterizing the
life-time of fluctuating Cooper pair and, in accordance to the uncertainty
principle, the inverse value of its characteristic energy scale $E_{0}\sim
k_{B}\left( T-T_{c0}\right)$.

One can see that Eq. (\ref{deltanu2}) is a sign-changing function and its
integral over the complete energy range must be equal zero:%
\begin{equation}
\int_{0}^{\infty }\delta \nu ^{\left( \mathrm{fl}\right) }\left( E,T\right)
dE=0.  \label{sumrule}
\end{equation}%
The statement (\ref{sumrule}) is nothing else as the sum rule:\
superconducting interactions cannot create new states, it just redistributes
existing ones to different energy levels. Namely, at the Fermi level a
sharp dip [$\delta \nu _{\left( 2\right) }^{\left( \mathrm{fl}\right)
}\left( 0,T\right) \sim -\mathrm{Gi}_{\left( 2\right)
}T_{c0}^{2}/(T-T_{c0})^{2}\nu _{n}$], the precursor of the superconducting gap is
formed, while the released states are moved to higher energies, with
maximum around $E_{0}\sim k_{B}\left( T-T_{c0}\right),$ the value
corresponding to the characteristic energy of fluctuation Cooper pairs (see
Fig.1a of the Main Text).

A major experimental tool for determining the density of states is by
measurements of the differential tunnel conductivity. Giaever and Megerle~%
\cite{GM61}, related the quasiparticle tunnel current to the densities of
electron states of the left and right electrodes and to the difference of
the equilibrium distribution functions in both of them (see Eq. (1) of the Main Text).
Assuming the left electrode being a normal metal with constant density
of states $\nu _{L}$ and the right electrode being a thin superconducting
film above its critical temperature one can write an explicit expression
for the excess tunnel conductivity in terms of $\delta \nu _{\left( 2\right)
}^{\left( \mathrm{fl}\right) }\left( E,T\right) $ and the derivative of the
Fermi function.
Combining the latter with the sum rule (\ref{sumrule}) one finds%
\begin{equation}
\delta \sigma _{\mathrm{tun}}^{\left( \mathrm{fl}\right) }\left( V\right)
\!=\!\frac{\hbar }{4TeR_{N}\nu _{n}}\!\int_{-\infty }^{\infty }\!\tanh
^{2}\left( \frac{E+eV}{2k_{B}T}\right) \delta \nu _{\left( 2\right)
}^{\left( \mathrm{fl}\right) }\left( E\right) \!dE\,  \label{tunfull}
\end{equation}%
and arrives at the disappointing conclusion that the predicted strong and
narrow singularity in the density of states Eq. (\ref{deltanu2}) manifests
itself in the observable tunnel conductivity only as a wide pseudogap structure ($eV_{\mathrm{%
pg}}\sim \Delta _{\mathrm{BCS}}$ instead of $E_{0}\sim k_{B}\left(
T-T_{c0}\right) $) and weak in the magnitude ($\ln \left( k_{B}T\tau
_{GL}/\hbar \right) \sim \ln \left[ T_{c0}/(T-T_{c0})\right] $ instead of $\
T_{c0}^{2}/(T-T_{c0})^{2}$), resembling that one in the
superconducting phase \cite{VD83} (see Fig.1b of the Main Text). Indeed,
due to the sum rule (\ref{sumrule}), almost the whole effect of fluctuations
on the tunnel current is averaged out in the process of energy integration.
The strong divergence of Eq. (\ref{deltanu2}) at zero energy is completely
eliminated due to presence of $\tanh ^{2}\left( E/2k_{B}T\right)$ in Eq. (%
\ref{tunfull}) and only a weak logarithmically singular behavior of the minimum and
two bumps of $\delta \sigma _{\mathrm{tun}}^{\left( \mathrm{fl}\right)
}\left( V\right)$ are reminiscent of the closeness to the superconducting transition.
The commonly accepted Giaver formula for the tunnel current does not allow to detect traces of the strong singularity of Eq. (\ref{deltanu2}%
), which should be manifested in the conductivity as a narrow zero bias anomaly in tunnel conductivity as we will see below.

\section{Where is the difference between the microscopic approach and Giaver
phenomenology hidden?}

One could be curious where does the difference between the microscopic
approach and the Giaver phenomenology lie? In order to understand this let
us follow the derivation of the latter from the former. Let us perform the
summation of the Green's functions of each electrode over the corresponding
momenta in Eq. (3) of the Main Text assuming the tunnel matrix elements to be momentum
independent. This makes the integrations of both Matsubara Green's functions
 independent and each of them can be presented in Lehmann form \cite%
{AGD}
\begin{equation}
\int \frac{d\mathbf{k}}{\left( 2\pi \right) ^{D}}G\left( \mathbf{k}%
,\varepsilon _{n}\right) =\int \frac{\nu \left( E\right) dE}{E-i\varepsilon
_{n}}.  \label{Lehmann}
\end{equation}%
Substituting Eq. (\ref{Lehmann}) into Eq. (3) of the Main Text, rewriting the
product of the energy denominators in the form of simple fractions, and summation over fermionic frequencies, gives%
\begin{eqnarray}
K\left( \omega _{\nu }\right) &=&\frac{1}{2\pi eR_{n}\nu _{L}\nu _{R}}\int
\int \frac{\nu _{L}\left( E_{L}\right) \nu _{R}\left( E_{R}\right)
dE_{L}dE_{R}}{E_{R}-E_{L}-i\omega _{\nu }}  \notag \\
&&\times \left[ \tanh \frac{E_{L}}{2T}-\tanh \left( \frac{E_{R}}{2T}-\frac{%
i\omega _{\nu }}{2\pi T}\right) \right] .  \label{Ktan}
\end{eqnarray}%
Looking at this expression one might be tempted to perform an analytic
continuation $i\omega _{\nu }\rightarrow \omega +i\delta $ ($\delta
\rightarrow 0$) and apply the Sokhotski--Plemelj theorem to the integration over $dE_{L}$ in Eq. (\ref{Ktan}):%
\begin{equation}
\lim_{\delta \rightarrow 0}\int_{a}^{b}\frac{\nu \left( E\right) }{E-i\delta
}dE=-i\pi \nu \left( 0\right) +\PVI\limits_{a }^{b}\frac{\nu \left( E\right)
}{E}dE,  \label{sokh}
\end{equation}%
where the ``dashed'' integral symbol means that the integral is performed in
the sense of a Cauchy principal value. 
This calculation of the imaginary part of Eq. (\ref{Ktan}) with subsequent use
of Eq. (2) of the Main Text immediately reproduces Giaever's and Megerle's formula, i.e.,
in accordance to the common believe, the microscopic approach confirms the
phenomenological result. 
Nevertheless, one should remember, that the
validity of Eq. (\ref{sokh}) requires the smoothness of the function $\nu
\left( E\right)$. 
However, this requirement is violated in the case under
consideration: as we saw above, the fluctuation correction $\delta \nu
_{\left( 2\right) }^{\left( \mathrm{fl}\right) }\left( E,T\right)$ close to the
transition temperature has a strong singularity at small energies. Hence,
performing the integration of the exact expression Eq.~(\ref{Ktan}) using the rule Eq. (\ref{sokh}), one looses the effect of the interplay
between the parameters $eV$ and $T-T_{c0}$ (or $\Delta _{\mathrm{BCS}}%
\widetilde{h}$ above the second critical field). 

The use of a finite-width $\delta$-function in the Sokhotski--Plemelj theorem washes out the result and makes the main difference.

\section{Model and Calculations}

We study the effect of SFs on the tunneling current $I\left(V\right) $
between a normal metal electrode and a disordered two-dimensional
superconducting film placed in a perpendicular magnetic field throughout the
whole phase diagram above the $H_{c2}(T)$ line. Describing this system by
means of a tunnel Hamiltonian, the tunnel-current can be expressed in terms
of the correlator $K\left( \omega _{\nu }\right) $ of the electron Green's
functions of the corresponding electrodes, which is analytically continued
from Matsubara frequencies $\omega _{\nu }=2\pi T\nu ,$ $\nu =0,1,2,...$ to
the upper half-plane of complex frequencies $\omega _{\nu}\rightarrow
-i\omega =-ieV$,~\cite{VD83,energy}:
\begin{equation}
I_{\mathrm{qp}}\left( V\right) =-e\Imm K^{R}(eV).  \label{IKgen}
\end{equation}

Being interested in low-transparency junctions and restricting our
consideration to the first order in $\mathrm{Gi}_{\left( 2\right) }$, one
can see that in the case the second electrode is not subject to
superconducting fluctuations -- e.g., is a normal STM tip -- the only
diagram which contributes to the tunnel-current is that presented in Fig. 2b of the Main Text.
 This diagram describes the suppression of the tunnel-current due
to the mechanism of fluctuation renormalization of the quasi-particle
density of states, discussed above.

In the absence of magnetic fields, the correlation function, Eq.~(\ref{IKgen}%
), was already studied in momentum representation~\cite{VD83}. The
generalization to the case of a perpendicular magnetic field can be made by
going over from the momentum to Landau representation with an appropriate
quantization of the Cooper pair motion (see, for example, Refs.~[%
\onlinecite{LV05, GVV11,glatz+epl11}]). Formally, this corresponds to a
replacement of the energy associated with the motion of the center of mass
of a free Cooper pair with momentum $\mathbf{q}$ by the eigen-energy of the
Landau state of level $m$: $\mathcal{D}\mathbf{q}^{2}$ $\rightarrow
\omega_c\left( m+1/2\right)$.
Here $\mathcal{D}$ is the electron diffusion coefficient and $\omega_c=4e%
\mathcal{D}H$ is the cyclotron frequency corresponding to the rotation of
the center of mass of a Cooper pair in a magnetic field $H$. The integration
over the two-dimensional momentum in correlator~(\ref{IKgen}) is replaced by
a summation over Landau levels according to the rule:
\begin{equation*}
\frac{\mathcal{D}}{8T}\int \frac{d^{2}q}{(2\pi )^{2}}f\left[ \mathcal{D}q^{2}%
\right] =\frac{h}{2\pi ^{2}t}\sum_{m=0}^{M}f\left[ \omega _{\mathrm{c}}(m+%
\frac{1}{2})\right] ,
\end{equation*}%
where $M=(T_{c0}\tau)^{-1}$ is a cut-off parameter related to the elastic
electron scattering time $\tau$ (see Ref.~[\onlinecite{GVV11}] for details).
This transformation is applied to the general expression for the correlation
function $K\left( \omega _{\nu }\right)$, and one finds \cite{VD83}:
\begin{align}
& K\left( \omega _{\nu }\right) =K^{\left( \mathrm{reg}\right) }\left(
\omega _{\nu }\right) +K^{\left( \mathrm{an}\right) }\left( \omega _{\nu
}\right)  \label{K1+K2} \\
& =\frac{2T_{c0}Sh}{\pi ^{3}\sigma _{n}R_{N}}\sum_{m=0}^{M}\left[
\sum_{k=0}^{\infty }+\sum_{k=-\nu }^{-1}\right] \frac{\left[ \mathcal{E}%
_{m}^{\prime }\left( k+2\nu \right) -\mathcal{E}_{m}^{\prime }\left(
k\right) \right] }{\mathcal{E}_{m}\left( |k|\right) }  \notag
\end{align}%
with $\sigma _{n}=e^{2}\nu _{n}\mathcal{D},$ $R_{N}$ being the tunneling
resistance of the junction and $S$ is its surface area. The function
\begin{equation}
\mathcal{E}_{m}\left( x\right) =\ln t+\psi \left[ \frac{1+x}{2}+\frac{4h}{%
\pi ^{2}t}\left( m+\frac{1}{2}\right) \right] -\psi \left( \frac{1}{2}\right)
\label{em}
\end{equation}%
represents the denominator of the fluctuation propagator (wavy line in Fig. 2b)
of the Main Text):
\begin{equation}
\mathcal{L}_{m}\left( x\right) =-\nu _{n}\mathcal{E}_{m}^{-1}\left( x\right)
,  \label{LE}
\end{equation}%
written in Landau representation and describing the fluctuation pairing of
electrons in the normal phase of a superconductor over a wide range of
temperatures and fields~\cite{LV05}. Here $t=T/T_{c0}$ and $h=\pi
^{2}/(8\gamma _{E}) H/H_{c2}(0)$ are dimensionless temperature and magnetic
field normalized by critical temperature and the value of second critical
field respectively, $\gamma _{E}=1.78$ is the exponential Euler constant.
The cyclotron frequency of a Cooper pair rotation in this parametrization is
$\omega _{\mathrm{c}}=\left(16hT_{c0}/\pi\right)$. We clarify that $\mathcal{%
E}_{m}^{\prime }\left( x\right)$ denotes derivative of the function $%
\mathcal{E}_{m}\left( x\right) $ with respect to its argument $x$,
explicitly given by
\begin{equation}
\mathcal{E}_{m}^{\prime }\left( x\right) =\frac{1}{2}\psi^{\prime } \left[
\frac{1+x}{2}+\frac{4h}{\pi ^{2}t}\left( m+\frac{1}{2}\right) \right] .
\label{deri}
\end{equation}

The two terms in Eq.~(\ref{K1+K2}) correspond to two fluctuation
contributions to the tunnel-current with different analytical properties.
Below we demonstrate how these contributions give rise to the pseudogap
maxima and the low-bias anomaly (LBA) in the tunneling conductivity in
two-dimensional disordered superconductors.

\subsection{Complete expression for the fluctuation tunnel-current}

We start our analysis with the first term of Eq.~(\ref{K1+K2}). Since the
external frequency $\omega _{\nu }$ enters the expression for $K^{(\mathrm{%
reg})}\left( \omega _{\nu }\right) $ only via the argument of the analytical
function $\mathcal{E}_{m}^{\prime }\left( k+2\nu \right) $ [see Eq.~(\ref%
{K1+K2})], one can easily perform its analytical continuation  by just
substituting $\omega _{\nu }\rightarrow -ieV$. Using Eq.~(\ref{IKgen}), one
finds for the general expression of the corresponding current $I^{(\mathrm{%
reg})}\left( V\right) $:
\begin{equation}
I^{(\mathrm{reg})}\left( V\right)\! =\!-\frac{2h}{\pi^{3}}\!\left( \frac {%
eT_{c0}S}{\sigma_{n}R_{N}}\right)\! \sum_{m=0}^{M}\sum_{k=0}^{\infty}\frac{%
\Imm\mathcal{E}_{m}^{^{\prime}}\left( k\!-\!\frac{ieV}{\pi T}\right) }{%
\mathcal{E}_{m}\left( k\right) }.  \label{I1}
\end{equation}

The second contribution to the tunneling current is determined by
\begin{equation}
K^{\left( \mathrm{an}\right) }\left( \omega _{\nu }\right) =\frac{2T_{c0}Sh}{%
\pi ^{3}\sigma _{n}R_{N}}\sum_{m=0}^{M}\sum_{k=1}^{\nu }f_{m}(k,\omega _{\nu
}),  \label{I2}
\end{equation}%
with%
\begin{equation}
f_{m}(k,\omega _{\nu })=\frac{\left[ \mathcal{E}_{m}^{\prime }\left( 2\nu
-k\right) -\mathcal{E}_{m}^{\prime }\left( k\right) \right] }{\mathcal{E}%
_{m}\left( k\right) }.  \label{fm}
\end{equation}%
Here the analytical contribution is more complex than in the case of $K^{(%
\mathrm{reg})}\left( V\right) ,$ since the frequency\ $\omega _{\nu }$\ is
not only present in the argument of function~(\ref{fm}) but also in the
upper limit of the sum over $k$ in Eq.~(\ref{I2}). Note, that this summation
limit can be reduced from $\nu $ to $\nu -1$ since $f_{m}(k=\nu ,\omega
_{\nu })=0$. The analytical continuation of a function of the form
\begin{equation*}
\theta _{m}\left( \omega _{\nu }\right) =\sum_{k=1}^{\nu -1}f_{m}(k,\omega
_{\nu })
\end{equation*}%
onto the upper half-plane of complex frequencies was performed in Ref.~[%
\onlinecite{AV80}] [see also Ref.~[\onlinecite{LV05}], equation (7.90)].
\begin{figure}[ptb]
\begin{center}
\includegraphics[ width=0.9\columnwidth ]{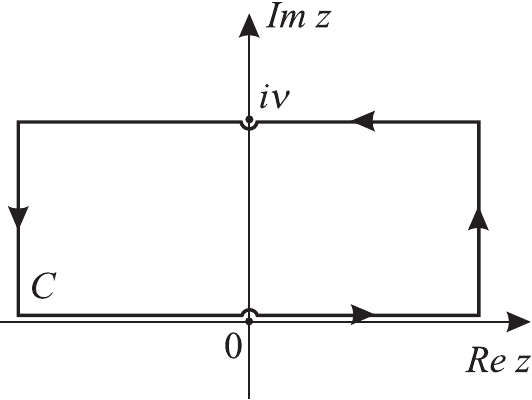}
\end{center}
\caption{Closed integration contour $\mathcal{C}$ in the plane of complex
frequencies.}
\label{mtcontour}
\end{figure}
By means of the Eliashberg transformation~\cite{E61} the corresponding sum
can be presented as a counterclockwise integral over a closed contour $%
\mathcal{C}$ consisting of two horizontal lines, two vertical lines, and two
semicircles in the upper complex plane, where the latter exclude the points $%
0$ and $i\nu$ (see Fig.~\ref{mtcontour}):
\begin{equation*}
\theta _{m}\left( \omega _{\nu }\right) \!=\!\frac{1}{2i}\ointctrclockwise%
\limits_{\mathcal{C}}\coth \left( \pi z\right) f_{m}(-iz,\omega _{\nu })dz\,.
\end{equation*}%
The integrals over the vertical line segments become zero, the integral over
the semi-circle at $z=i\nu $ is zero since $f_{m}(k=\nu ,\omega _{\nu })=0$,
the integral over the semi-circle at $z=0$ reduces to the residual of $\coth
\left( \pi z\right) $. Inverting the direction of integration over the line
segment with $\Imm z=\nu $ and then shifting the integration variable as $%
z+i\omega _{\nu }/2\pi T\rightarrow z_{1}$ in the corresponding integral,
one finds:
\begin{eqnarray}
& \theta_{m}\left( \omega_{\nu}\right) \!=-\!\frac{f_{m}(0,\omega_{\nu})}{2}+%
\frac{1}{2i}\PVI_{-\infty}^{\infty}\coth\left( \pi z\right)  \notag \\
& \times\left[ f_{m}(\!-iz,\omega_{\nu})\!-\!f_{m}(\!-iz\!-\!\omega_{\nu}/2%
\pi T,\omega_{\nu})\right] dz.  \label{resi}
\end{eqnarray}
 Eq.~(\ref{resi}) is already an
analytical function of $\omega _{\nu }$ and one can perform its continuation
just by the standard substitution $\omega _{\nu }\rightarrow -i\omega$.
Shifting the variable in the second integral again as $z-\omega /2\pi
T\rightarrow z_{2}$ and using the identity
\begin{equation*}
\coth a-\coth b=-\frac{\sinh \left( a-b\right) }{\sinh a\sinh b}
\end{equation*}%
one finally finds%
\begin{align}
& \theta _{m}^{R}\left( -i\omega \right) \!=-\!\frac{f_{m}(0,-i\omega )}{2}
\label{thetar} \\
& -i\frac{\sinh \left( \omega /2T\right) }{2}\PVI\limits_{-\infty }^{\infty }%
\frac{f_{m}(-iz,-i\omega )dz\!}{\sinh \left( \pi z\right) \sinh \pi \left(
z+\omega /2\pi T\right) }.  \notag
\end{align}%
Substituting the explicit expression for function $f_{m}(-iz,-i\omega )$
from Eq.~(\ref{fm}) into Eq.~(\ref{thetar}) results in
\begin{align}
& K^{\left( \mathrm{an}\right) R}\left( -i\omega \right) =-\frac{T_{c0}Sh}{%
\pi ^{3}\sigma _{n}R_{N}}\sum_{m=0}^{M}\left\{ \frac{\left[ \mathcal{E}%
_{m}^{\prime }\left( -\frac{i\omega }{\pi T} \right) -\mathcal{E}%
_{m}^{\prime }\left( 0\right) \right] }{\mathcal{E}_{m}\left( 0\right) }%
\right.  \label{KR2} \\
& \left. +i\sinh \left( \frac{\omega }{2T}\right) \PVI\limits_{-\infty
}^{\infty }\frac{\left[ \mathcal{E}_{m}^{\prime }\left( iz\!-\!\frac{i\omega
}{\pi T}\right)\!-\! \mathcal{E}_{m}^{\prime }\left(\!-iz\right) \right] dz\!%
}{\mathcal{E}_{m}\left( -iz\right) \sinh \left( \pi z\right) \sinh \pi
\left( z+\frac{\omega} {2\pi T}\right) }\right\} .  \notag
\end{align}

Eqs. (\ref{IKgen}) and (\ref{KR2}) determine the second fluctuation
contribution to the tunneling current $I^{\left( \mathrm{an}\right) }\left(
V\right)$.

Let us note that the first term of $K^{\left( \mathrm{an}\right) R}$ is
nothing but half of the first summand (with $k=0$) of the sum in Eq.~(\ref%
{I1}) with opposite sign. Technically it would be easy to incorporate the
latter into $K^{\left( \mathrm{reg}\right) R}$. However, such a procedure
would be physically misleading: we will see below that this $k$ and $z$
independent term in Eq.~(\ref{KR2}) cancels the corresponding linear
contribution stemming from the integral term at small voltages. As a result,
the current $I^{\left( \mathrm{an}\right) }\left( V\right) $, determined by
the imaginary part of Eqs.~(\ref{KR2}), does not contain a linear
contribution if expanded in powers of voltage. This means that it does not
contribute to the magnitude of the differential tunnel conductivity at zero
voltage $\sigma _{\mathrm{tun}}^{\left( \mathrm{fl}\right) }\left(
T,H,V=0\right) =dI^{\left( {\mathrm{fl}}\right) }/dV|_{V=0}$, which is the
easiest quantity to measure in experiments. Nevertheless, it contributes to
the current-voltage characteristics at finite voltages and, as we will see
below, can noticeably manifest itself even at very low voltages $eV\sim
T-T_{c0}$ as a LBA.

Adding $I^{\left( \mathrm{reg}\right) }\left( V\right) $ and $I^{\left(
\mathrm{an}\right) }\left( V\right) $ one finds the general expression for
the fluctuation contribution to the tunnel-current, which is valid in the
complete phase diagram beyond the $H_{c2}\left( T\right) $ line:
\begin{widetext}%
\begin{align}
&I^{\left(  \mathrm{fl}\right)  }\left(  t,h,V\right)  =I^{\left(
\mathrm{reg}\right)  }+I^{\left(  \mathrm{an}\right)  }=-\frac{2eT_{c0}Sh}%
{\pi^{3}\sigma_{n}R_{N}}\sum_{m=0}^{M}\sum_{k=0}^{\infty}\frac
{\Imm\Em_{m}^{\prime}\left(  k-ieV/\pi T\right)
}{\Em_{m}\left(  k\right)  }+\frac{eT_{c0}Sh}{\pi^{3}\sigma_{n}R_{N}%
}\sum_{m=0}^{M}\left\{  \frac{\Imm\Em_{m}^{\prime}\left(
-ieV/\pi T\right)  }{\Em_{m}\left(  0\right)  }\right.
\label{Itot}\\
&  \left.  +\sinh\left(  \frac{eV}{2T}\right)  \PVI\limits_{-\infty
}^{\infty}dz\frac{\operatorname{Re}\Em_{m}\left(  iz\right)  \left[
\operatorname{Re}\Em_{m}^{\prime}\left(  iz-ieV\right)
-\operatorname{Re}\Em_{m}^{\prime}\left(  iz\right)  \right]
+\Imm\Em_{m}\left(  iz\right)  \left[  \Imm%
\Em_{m}^{\prime}\left(  iz-ieV\right)  +\Imm%
\Em_{m}^{\prime}\left(  iz\right)  \right]  }{\sinh\left(  \pi
z\right)  \sinh\left[  \pi\left(  z-eV/2\pi T\right)  \right]  \left[
\Rem^{2}\Em_{m}\left(  iz\right)  +\Imm%
^{2}\Em_{m}\left(  iz\right)  \right]  }\right\}  .\nonumber
\end{align}
\end{widetext}

Eq. (\ref{Itot}) is the main result of this work. The first term $I^{\left(
\mathrm{reg}\right) }\left( V\right)$ has been studied in detail for
different limiting cases using different approaches: close to $T_{c0}$,~\cite%
{VD83, CCRV90,V93,L10}: (i) in a wide temperature range in zero field~\cite%
{VD83}, or (ii) close to $T_{c0}$ in magnetic fields $H\ll H_{c2}(0)$,~\cite%
{R93}. The current contribution $I^{\left( \mathrm{an}\right) }\left(
V\right) $ has been omitted in all these works based on the ``standard''
argument that the zero frequency bosonic mode (which traverses through the
propagator) is singular in the vicinity of the transition. However, it is
known that this argument sometimes works (e.g., in the case of the
Maki-Thompson contribution to conductivity~\cite{M68}), but also sometimes
fails (e.g., for the Aslamazov-Larkin contribution to conductivity~\cite%
{AL68}). In our case this argument turns to out be correct only for very
small voltages. The reason being that voltage itself, together with
temperature deviations from the transition point and finite magnetic fields,
drives the system away from the immediate vicinity of the transition, which
invalidates the argument regarding the dominance of the zero frequency
bosonic mode.

\begin{figure*}[tbh]
\begin{center}
\includegraphics[ width=0.44\textwidth ]{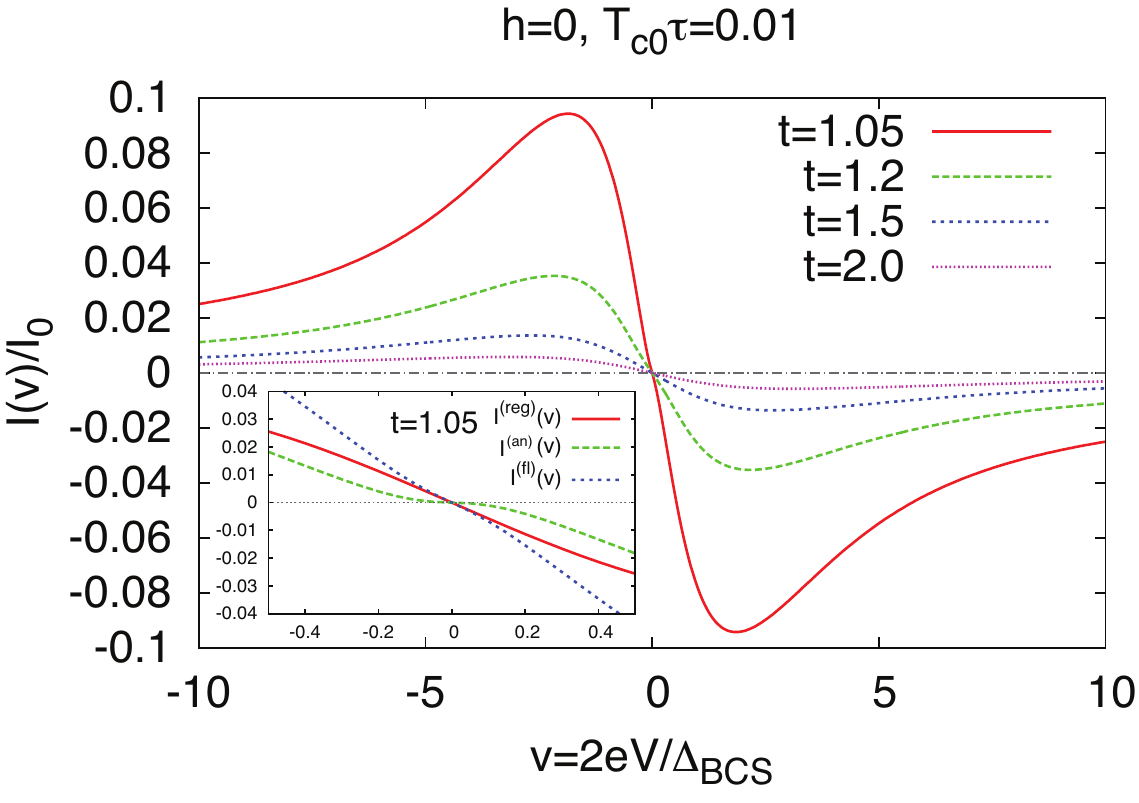}
\includegraphics[
width=0.44\textwidth ]{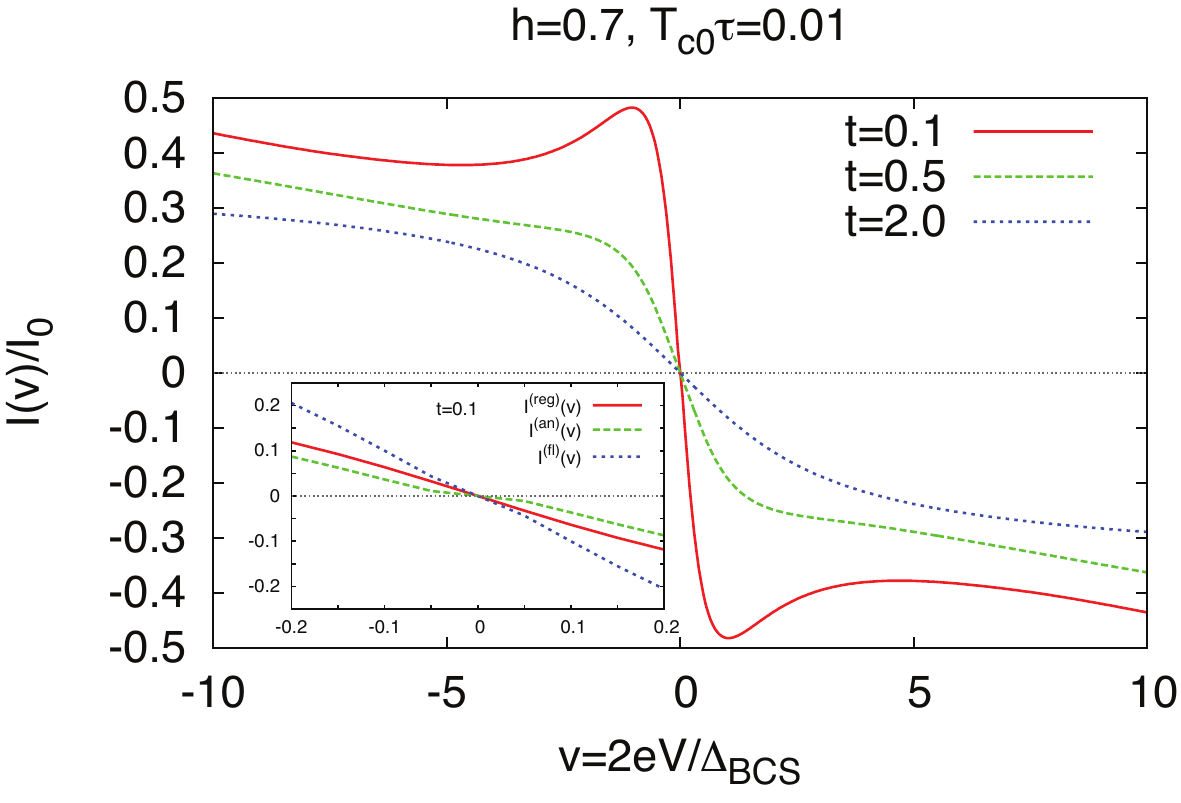}
\end{center}
\caption{(Color online) Total tunneling current close to $T_{c0}$ (left) and
near $h_{c2}(0)$ (right) at various temperatures depending on the
dimensionless voltage $v=2eV/\Delta_{\mathrm{BCS}}$. The insets show the
regular and anomalous contributions at the respective lowest temperature
separately. As one can see, the anomalous part has a nonlinear component
near $v=0$. The current is normalized to $I_0=eT_{c0}S/(\protect\sigma_n
R_N) $. (left) parameter points in Fig.~\ref{fig.plots} are $d_1$-$%
d_4$, inset $d_1$, (right) parameter points are $g_1$-$g_3$, inset $g_1$.}
\label{fig.tunnelIV}
\end{figure*}

We present several plots of the tunnel-current and the tunnel conductance.
Since they depend on three parameters: $t$, $h$, and $v$, only lines or
planes in the full parameter space are presented as line or surface plots.
In Fig. 3 of the Main text all parameter points and lines
in the $t$-$h$ phase diagram for all following figures are shown. The critical field
line, $h_{c2}(t)$, separating the superconducting (SC) and the normal
fluctuation region (SF) is defined by $\mathcal{E}_0(0)=0$. Each figure
caption refers to these parameter locations. Fig.~\ref{fig.tunnelIV} shows
the behavior of $I^{\left( \mathrm{fl}\right) }\left( t,h,V\right)$ near $%
T_{c0}$ and $H_{c2}(0)$.

In the following we will carefully analyze the effect of superconducting
fluctuations in the whole phase diagram. We start our discussion with the
regular contribution $I^{\left( \mathrm{reg}\right) }\left( V\right) $ and
then elucidate the important role of the anomalous contribution $I^{\left(
\mathrm{an}\right) }\left( V\right)$, which was neglected in literature so
far.

\subsection{Analysis of the asymptotic behavior of $I^{\left( \mathrm{reg}%
\right) }\left( V\right) $}

\emph{Close to $T_{c0}$ and for sufficiently weak magnetic fields $H\ll
H_{c2}\left( 0\right) $}, the most singular term in Eq. (\ref{I1}) arises
from the zero frequency bosonic mode $k=0$, when the propagator has a pole
at $\epsilon=0$ and
\begin{equation}
\mathcal{E}_{m}\left( 0\right) =\epsilon +2h\left( m+\frac{1}{2}\right)
\label{em0}
\end{equation}%
with $\epsilon =\ln t\approx t-1\ll 1$ as reduced temperature. The summation
over Landau levels can be performed in terms of polygamma-functions, $%
\psi^{(n)}(x)$, and one finds an expression valid for any combination of $%
\epsilon$ and $h\ll 1$:
\begin{align}
I^{(\mathrm{reg})}\left( V,h,\epsilon \right) =& -\frac{eTS}{2\pi ^{3}\sigma
_{n}R_{N}}\left[ \ln \frac{1}{2h}-\psi \left( \frac{1}{2}+\frac{\epsilon }{2h%
}\right) \right]  \notag \\
& \cdot \Imm\psi ^{\prime }\left( \frac{1}{2}-\frac{ieV}{2\pi T}\right).
\label{eq.Ireg}
\end{align}%
Eq.~(\ref{eq.Ireg}) reproduces the results of Refs. ~[\onlinecite{VD83,R93}%
]. The corresponding contribution to the tunneling conductance is
\begin{align}
\sigma ^{(\mathrm{reg})}\left( V\right) & =\frac{Se^{2}}{4\pi ^{4}\sigma
_{n}R_{N}}\left[ \ln \frac{1}{2h}-\psi \left( \frac{1}{2}+\frac{\epsilon}{2h}%
\right) \right]  \notag \\
& \cdot \Rem\psi ^{\prime \prime }\left( \frac{1}{2}-\frac{ieV}{2\pi T}%
\right) .
\end{align}

\emph{In the region of high temperatures $T\gg T_{c0}$ and zero magnetic
field} we restrict our analytical consideration to the fluctuation
contribution to the differential conductivity at zero voltage. Performing an
integration instead of a summation in Eq.~(\ref{I1}) one finds%
\begin{equation*}
\sigma ^{(\mathrm{reg})}(0,t\gg 1)=-\frac{Se^{2}}{4\pi ^{2}\sigma _{n}R_{N}}%
\left( \ln \frac{\ln \frac{1}{T_{c0}\tau }}{\ln t}\right) ,
\end{equation*}%
which is again in complete agreement with Ref.~[\onlinecite{VD83}].

\emph{Close to the line $H_{c2}\left( t\right) $ and for sufficiently low
temperatures $t\ll h_{c2}(t)$} the lowest Landau level approximation (LLL)
holds. The corresponding propagator (with quantum number $m=0)$ has a pole
structure and Eq.~(\ref{em}) acquires the form:
\begin{equation}
\mathcal{E}_{0}\left( k\right) =\widetilde{h}+\frac{\pi ^{2}tk}{4h_{c2}}
\label{eline}
\end{equation}%
with $\widetilde{h}\left( t\right) =\left( H-H_{c2}\left( t\right) \right)
/H_{c2}\left( t\right)$. Keeping only the $m=0$ term in Eq. (\ref{I1}), one
can write
\begin{equation}
I^{(\mathrm{reg})}\left[V,t\!\ll\! h_{c2}(t)\right]\!=\!-\frac{2eT_{c0}Sh}{%
\pi ^{3}\sigma _{n}R_{N}}\sum_{k=0}^{\infty }\frac{\Imm\mathcal{E}%
_{0}^{\prime }\left( k-\frac{ieV}{\pi T}\right) }{\widetilde{h}+\frac{\pi
^{2}tk}{4h_{c2}(t)} }.  \label{sum4}
\end{equation}%
The imaginary part $\Imm\mathcal{E}_{0}^{\prime }\left( k-ieV/\pi T\right)$
can be explicitly written using Eq.~(\ref{deri}) in the limit $t\ll
h_{c2}(t) $ and the asymptotic behavior of $\psi ^{\prime }\left( |x|\gg
1\right)\sim 1/x$:
\begin{equation}
\Imm\mathcal{E}_{0}^{\prime }\left( k-\frac{ieV}{\pi T}\right) =\frac{eV}{%
2\pi T}\frac{1}{\left[ k+\frac{4h_{c2}(t)}{\pi ^{2}t}\right] ^{2}+\left(
\frac{eV}{\pi T}\right) ^{2}}.
\end{equation}%
The summation in Eq.~(\ref{sum4}) can then be performed exactly in terms of
polygamma-functions, i.e., using
\begin{equation*}
\sum_{k=0}^{\infty }\frac{1}{k+\alpha }\frac{1}{\left( k+\beta \right)
^{2}+\gamma ^{2}}=\frac{1}{\gamma }\Imm\frac{\psi \left( \alpha \right)
-\psi \left( \beta +i\gamma \right) }{\beta +i\gamma -\alpha }\,,
\end{equation*}%
which gives an expression for the regular part of the fluctuation current
valid for low enough temperatures along the line $h_{c2}(t)$:
\begin{widetext}
\begin{equation}
I^{\left( \mathrm{reg}\right) }\left[ v_{t},t\ll h_{c2}\left( t\right) %
\right] =-\frac{2eST_{c0}h}{\pi ^{3}\sigma _{n}R_{N}}\frac{v_{t}}{1+v_{t}^{2}%
}\left\{ \left[ \ln \left( \frac{4h_{c2}\left( t\right) }{\pi ^{2}t}\right)
\sqrt{1+v_{t}^{2}}\allowbreak \allowbreak -\psi \left( \frac{4h_{c2}}{\pi
^{2}t}\widetilde{h}\right) \right] -\frac{\arctan v_{t}}{v_{t}}\right\}.
\label{curr1a}
\end{equation}%
\end{widetext}

Here, we introduced the dimensionless voltage
\begin{equation*}
v_{t}=\frac{\pi eV}{4h_{c2}\left( t\right) T_{c0}},
\end{equation*}
which defines the characteristic scale of $\sigma^{\left( \mathrm{reg}%
\right)}$ in the considered domain of the phase diagram. We stress, that
this scale depends on temperature via the parameter $h_{c2}\left( t\right)$.

\emph{Close to $H_{c2}\left( 0\right)$, in the region of very low
temperatures $t\ll \widetilde{h}$}, the argument of the $\psi$-function in
Eq.~(\ref{curr1a}) becomes large despite the smallness of $\widetilde{h}$,
and the $\psi$-function can therefore be approximated by its asymptotic
expression. One gets
\begin{align}
I^{(\mathrm{reg})}\left( v,t\ll \widetilde{h}\right) = -\frac{eS\Delta _{%
\mathrm{BCS}}}{4\pi ^{2}\sigma _{n}R_{N}}  \notag \\
\cdot \frac{v}{1+v^{2}}\left[ \ln \frac{\sqrt{1+v^{2}}}{\widetilde{h}}-\frac{%
\arctan v}{v}\right]  \label{QF}
\end{align}%
with $\Delta _{\mathrm{BCS}}=\pi T_{c0}/\gamma _{E}$ being the value of BCS\
gap. The characteristic scale where the maximum of the tunnel conductance
appears at these low temperatures is $v=2eV/\Delta _{\mathrm{BCS}}\sim 1$,
i.e.
\begin{equation}
eV_{\max }\sim \Delta _{\mathrm{BCS}}.  \label{vmax}
\end{equation}

\emph{In the region of high fields $H\gg H_{c2}$ and low temperatures}, the
asymptotic behavior of the tunneling current can be studied in complete
analogy to the case of high temperatures and weak fields. The sums in Eq. (%
\ref{I1}) can be approximated by integrals, which gives for the value of the
differential conductivity at zero voltage: 
\begin{equation*}
\sigma ^{(\mathrm{reg})}\left( 0,h\gg 1\right) =-\frac{e^{2}S}{4\pi
^{2}\sigma _{n}R_{N}}\left( \ln \frac{\ln \frac{1}{T_{c0}\tau }}{\ln h}%
\right)\,.
\end{equation*}%
One can see that this dependence is exactly the same as that one in the case
of high temperatures with reversed roles of the reduced temperature and the
reduced field.

\subsection{Low voltage behavior of $I^{\left( \mathrm{an}\right) }\left(
V\right) $}

\begin{figure}[tbh]
\begin{center}
\includegraphics[width=0.99\columnwidth ]{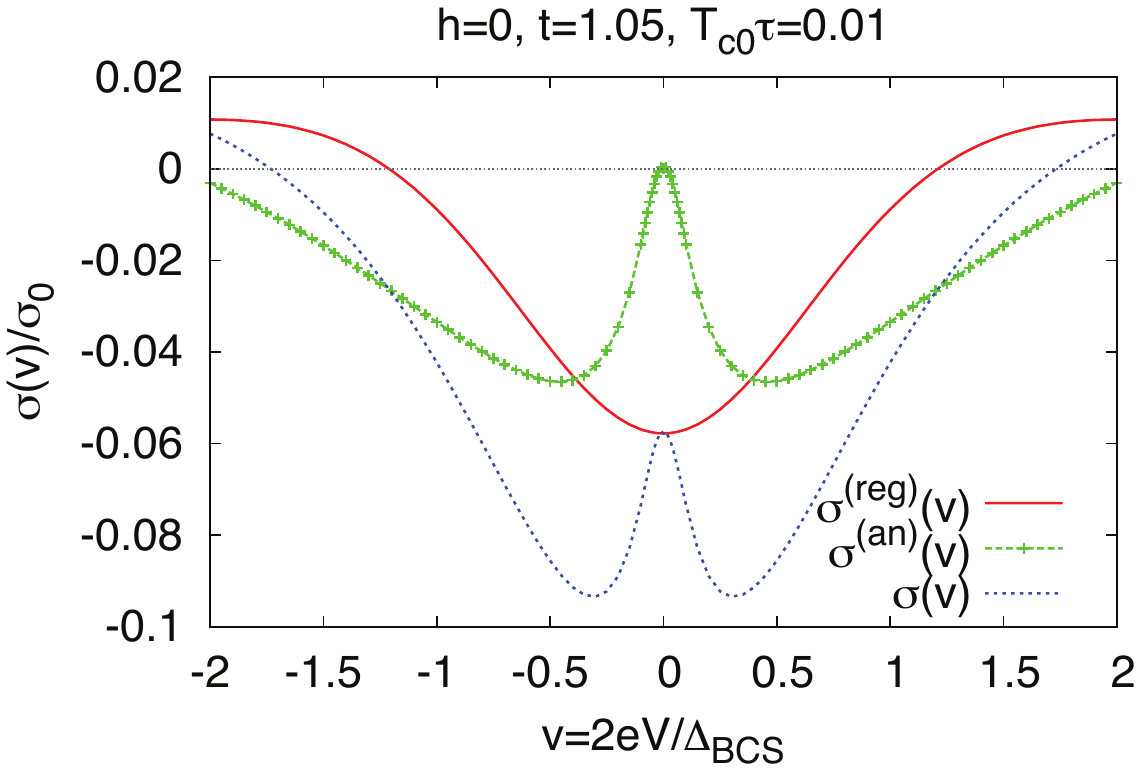} %
\includegraphics[width=0.99\columnwidth ]{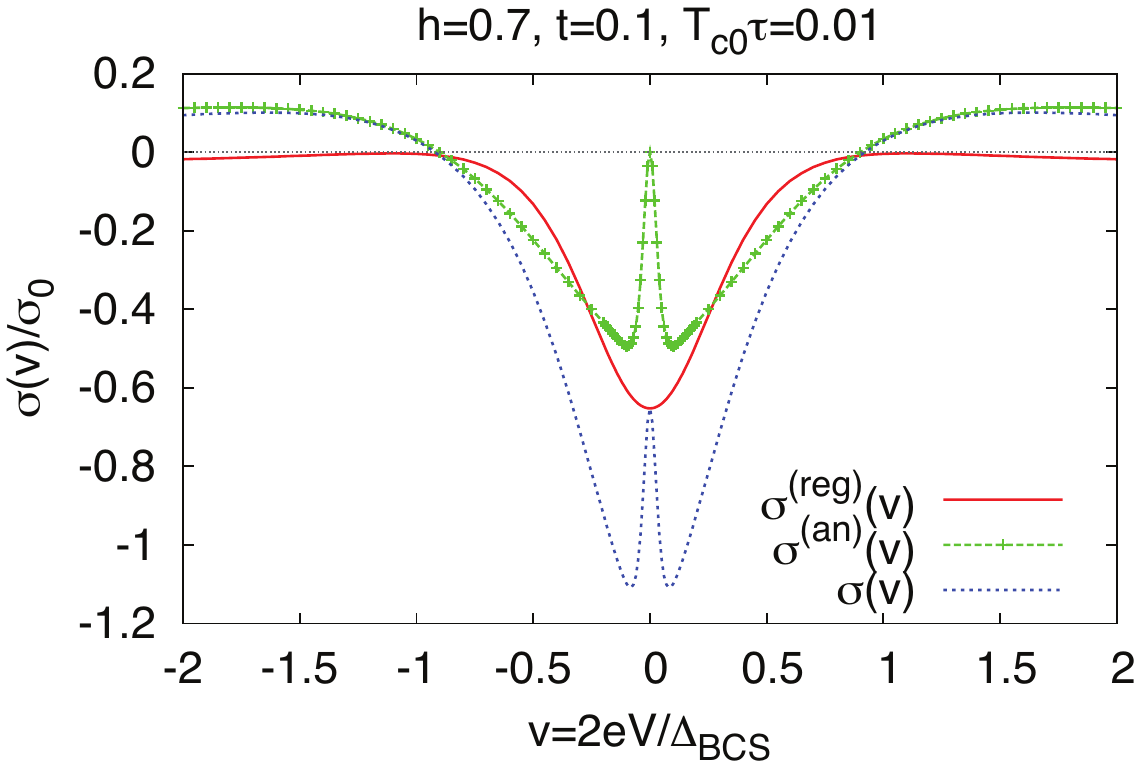}
\end{center}
\caption{Regular and anomalous contributions to the tunneling
conductance close to $T_{c0}$ (top) and at low temperatures near $h_{c2}(0)$
(bottom). The regular part is presented by a solid line (red), the anomalous
by crossed line (green) and the their sum, i.e., the total fluctuation
contribution, is shown by a dashed line (blue). (top) parameter point in
Fig.~\ref{fig.plots}  is $d_{1}$ [see also Fig.~\ref{fig.plots} c)], (bottom) $g_{1}$.}
\label{tuncond_reg_an}
\end{figure}

In the low-voltage limit, $V\rightarrow 0$, the general expression for $%
I^{\left( \mathrm{an}\right) }\left( V\right)$, (\ref{Itot}), can be
expanded in small $eV$. We start with the first order term of that
expansion, where one can assume $V=0$ in the argument of integrand function
and obtain

\begin{eqnarray}
&&I^{\left( \mathrm{an}\right) }\left( V\rightarrow0\right)\! =\!\frac{%
eT_{c0}Sh}{\pi^{3}\sigma_{n}R_{N}}\sum_{m=0}^{M}\left\{ \frac {\Imm\mathcal{E%
}_{m}^{\prime}\left( \! -ieV/\pi T\right) }{\mathcal{E}_{m}\left( 0\right) }%
\right.\!+  \notag \\
&&\!\!\left.\frac{eV}{T}\int_{-\infty}^{\infty }dz\frac{\Imm\mathcal{E}%
_{m}\left( iz\right) \Imm\mathcal{E}_{m}^{\prime}\left( iz\right) }{%
\sinh^{2}\pi z\left[ \Rem^{2}\mathcal{E}_{m}\left( iz\right) \! +\!\Imm^{2}%
\mathcal{E}_{m}\left( iz\right) \right] }\right\}.  \label{ian}
\end{eqnarray}

In the region of temperatures close to the transition temperature $T_{c0}$
and along the transition line for temperatures $t\ll h_{c2}(t)$, the
propagator has a simple pole structure [see Eqs.~(\ref{em0}) and (\ref{eline}%
)] and the integral in Eq.~(\ref{ian}) can be calculated analytically.
Performing this integration one finds that the second term of Eq.~(\ref{ian}%
) exactly annihilates the linear part of the first term. This fact justifies
the static approximation (zero frequency bosonic mode) made in Refs.~[%
\onlinecite{VD83,CCRV90,V93,R93,L10}]. Yet, this static approximation turns
out to be valid only for very low voltages. Expanding the integrand in Eq.~(%
\ref{Itot}) to higher orders in voltage reveals an unexpected result. One
can see that the voltage $V$ enters the integrand of Eq.~(\ref{Itot}) in two
different places: in the argument of $\mathcal{E}_{m}\left( iz-ieV\right) $
in the numerator and in the argument of $\sinh \left( \pi z-eV/2T\right) $
in the denominator. The expansion of $\mathcal{E}_{m}\left( iz-ieV\right) $
results in the appearance of a weakly voltage-dependent term of the order of
$O\left( V^{3}/T_{c0}^{3}\right) $ in $I^{\left( \mathrm{an}\right) }$,
while, as one can easily verify, the expansion of $\sinh ^{-1}\left( \pi
z-eV/2T\right) $ up to the third order in voltage after integration leads to
a very singular correction%
\begin{eqnarray}
I^{\left( \mathrm{an}\right) }\left( \epsilon ,V\right) &=&-\frac{\pi
e^{2}V^{3}\left\vert \psi ^{\prime \prime }\left( \frac{1}{2}\right)
\right\vert e^{2}S}{2^{8}\pi ^{4}\sigma _{n}R_{N}T^{2}}  \label{ancur} \\
&&\cdot \int_{0}^{\infty }dy\int_{\gamma }^{\infty }\frac{dz}{z^{2}\left[
\left( \epsilon +y\right) ^{2}+\left( z\right) ^{2}\right] }.  \notag
\end{eqnarray}

The strong divergency of this expression at small frequencies indicates that the process of generating current $\ $(\ref{ancur}) should be
limited in time. Indeed, from the physical picture described above, it is clear that the processes of anomalous Cooper pairings of the
injected electrons take place until the latter remain non-thermalized., i e.
for times shorter than $\tau _{\phi }.$ Hence the frequency integral should
be cut-off at $\omega \sim \tau _{\phi }^{-1},$ what in dimensionless
variables corresponds $z_{\min }=$ $\gamma =\frac{\pi ^{2}}{8T_{c}\tau _{e}}$. 
Further integration is trivial and one finds for the non-linear current the
expression%
\begin{eqnarray*}
I^{\left( \mathrm{an}\right) }\left( \epsilon ,V\right)  &=&-\frac{%
7e^{2}S\zeta \left( 3\right) }{2^{8}T^{2}\pi ^{3}\sigma _{n}R_{N}}\frac{%
e^{2}V^{3}}{\epsilon \gamma } \\
&&\cdot \left\{ 1-\left( \frac{\gamma }{\epsilon }\right) \arctan \frac{%
\epsilon }{\gamma }+\left( \frac{\epsilon }{\gamma }\right) \arctan \frac{%
\gamma }{\epsilon }\right\}
\end{eqnarray*}%
which valid for $\gamma ,\epsilon \ll 1.$

The corresponding contribution to the differential conductivity is

\begin{eqnarray}
\sigma _{\mathrm{tun}}^{\left( \mathrm{an}\right) }\left( \epsilon ,V\right)
&=&-\frac{21\zeta \left( 3\right) e^{2}S}{2^{8}\pi ^{3}\sigma _{n}R_{N}}%
\frac{e^{2}V^{2}}{T^{2}\epsilon \gamma }  \label{difan} \\
&&\cdot \left\{ 1-\left( \frac{\gamma }{\epsilon }\right) \arctan \frac{%
\epsilon }{\gamma }+\left( \frac{\epsilon }{\gamma }\right) \arctan \frac{%
\gamma }{\epsilon }\right\}  \notag
\end{eqnarray}

Eq. (\ref{difan}) describes two different regimes. The first corresponds to the
growth of the LBA when temperature approaches $T_{c0}$ but $T-T_{c0}$
remains larger than the inverse energy relaxation time $\tau _{\phi }^{-1}:$%
\begin{equation*}
\sigma _{\mathrm{tun}}^{\left( \mathrm{an}\right) }\left( eV\ll T\epsilon
\right) =-\frac{21\zeta \left( 3\right) e^{2}S\tau _{\phi }}{2^{4}\pi
^{5}\sigma _{n}R_{N}}\frac{e^{2}V^{2}}{T-T_{c0}}.
\end{equation*}%
Notoriously that the magnitude is directly proportional to the border area
volue where the energy relaxation of the injected electrons takes place.
When $T-T_{c0}$ reaches the value of $\tau _{\phi }^{-1}$ the LBA is
saturated and does not grow more:%
\begin{equation*}
\sigma _{\mathrm{tun}}^{\left( \mathrm{an}\right) }\left( \epsilon ,V\right)
=-\frac{21\zeta \left( 3\right) e^{2}S}{2^{9}\pi ^{2}\sigma _{n}R_{N}}\frac{%
e^{2}V^{2}}{T^{2}\gamma ^{2}}.
\end{equation*}%
The complete expression for small voltages ($eV\ll T\epsilon $) and in the
case of low energy relaxation ($\gamma \ll \epsilon $) is%
\begin{equation*}
\sigma _{\mathrm{tun}}^{\left( \mathrm{an}\right) }\left( \epsilon ,V\right)
=-\frac{7\zeta \left( 3\right) e^{2}S}{2\pi ^{4}\sigma _{n}R_{N}}\left[ \ln
\frac{T_{c0}}{T-T_{c0}}-\frac{3\tau _{\phi }}{8\pi }\frac{e^{2}V^{2}}{%
T-T_{c0}}\right] .
\end{equation*}%
From this expression one can estimate for the width of the peak:

\begin{equation*}
eV_{\mathrm{LBA}}\sim \sqrt{\frac{T-T_{c0}}{\tau _{\phi }}}\ln ^{1/2}\frac{%
T_{c0}}{T-T_{c0}}.
\end{equation*}%
In the case of strong energy relaxation the anomalous contribution becomes
of the order of the higher contributions of the regular part and it is not
observable on the background of the pseudogap structure.

The effect of both fluctuation contributions, $I^{\left( \mathrm{reg}\right)
}\left( V\right) $ and $I^{\left( \mathrm{an}\right) }\left( V\right)$, on
the tunneling conductance is demonstrated in Fig.~\ref{tuncond_reg_an}.
Similar behavior can be observed along the whole line $H_{c2}\left( T\right)
.$\ The singularity in the low voltage behavior of tunneling conductance
rapidly smears out when moving away from the transition line or increasing
the temperature [see Fig.~\ref{fig.plots}].

\subsection{Numerical analysis}

The temperature, magnetic field, and voltage dependencies of the tunneling
conductance due to superconducting fluctuations, calculated numerically
based on Eq.~(\ref{Itot}), are presented in Figs.~3a,f,h) as
surface plots. The numerical procedure to calculate the $k$-sum of the first
term needs to take into account its relatively slow convergence. Therefore
it is calculated explicitly up to a threshold at which the sum can be
replaced by an integral and the polygamma functions by their asymptotic
behavior. (here we use as threshold-$k$, the value $k_M$ at which the
argument of the function $\mathcal{E}_m$ reaches $1000$). The
``rest''-integrals are calculated with inverse integration variable using a
Gauss-Legendre method. The second term requires a careful treatment of the
two integrable poles, which is done by analytical calculation of the
residuals in a small interval around them, where the denominator is
linearized. Also the numerical integration outside the pole intervals is
done by using adaptive integration point distances. The overall behavior of
both terms of the tunnel-current results in a pronounced pseudo-gap
structure of the conductance near the superconducting region. It is the
non-linear anomalous term of the tunnel-current which is responsible for the
fine structure (``local maximum'') at the center of the gap, the LBA.

At this point it is worth mentioning that another sharp fine structure of
tunnel conductance which should occur in the same scale $eV \sim T-T_{c0}$
was predicted in Ref.~[\onlinecite{VD83}]. This structure appears due to
\textit{interaction of fluctuations} as the second order correction in
Ginzburg-Levanyuk number $\mathrm{Gi}_{\left(2\right)}$ (but still in first
order in the barrier transparency). This contribution has an interference
nature (analogously to Maki-Thompson process) and, in contrast to the
discussed above nonlinear contribution $\sigma^{\left(\mathrm{an}\right)
}\left( eV \ll T-T_{c0} \right)\sim \mathrm{Gi}_{\left(2\right)}
[eV/(T-T_{c0})]^2$, diverges at zero voltage as $\mathrm{Gi}%
^2_{\left(2\right)}[T_{c0}/(T-T_{c0})]^2\ln [(T-T_{c0})/eV]$. Such
divergency, in complete analogy to Maki-Thompson contribution, is cut off by
any phase-breaking mechanism~\cite{M68,T70}.

Analyzing the surface plot representation of the experimental results of
Ref.~[\onlinecite{Benjamen10}], obtained at temperature close to $T_{c0}$,
one notices their striking similarity to the theoretical surfaces presented
in Fig.~\ref{3DT}. Indeed, the authors of Ref.~[\onlinecite{Benjamen10}]
mentioned the agreement of their results with the theoretical prediction of
Ref.~[\onlinecite{VD83}]. Fig.~\ref{3DH} shows how the corresponding surface
transforms at low temperatures and strong magnetic fields close to $%
H_{c2}(0) $. 

It is interesting to note that the behavior of the general expression~(\ref%
{Itot}) clearly shows growth of the fluctuation effects in the domain of
intermediate temperatures and magnetic fields, beyond the immediate vicinity
of $T_{c0}$ and $H_{c2}(0)$, see plots of the zero-bias tunnel conductance $%
\sigma(t,h,0)=\sigma^{\mathrm{(reg)}}(t,h,0)$ in Figs.~\ref{fig.plots}b,e) . In Fig.~\ref{fig.plots}f) one can see the evolution of
the pseudogap near the $h_{c2}(t)$ line (slightly offset by a factor $1.1$,
see caption), exhibiting a deeper suppression for intermediate temperatures
and fields. This fact is in agreement with the general ideas of the theory
of fluctuations establishing the growth of fluctuations strength
(characterized by the Ginzburg-Levanyuk number) as one moves away from the
extreme points [$T_{c0}$ and $H_{c2}(0)$] of the curve $H_{c2}(T)$ (see
chapter 2 of Ref.~[\onlinecite{LV05}]).


\begin{thebibliography}{99}
\bibitem{energy} Below we will use the convention $\hbar =k_{\mathrm{B}}=c=1$%
.
\end{thebibliography}
\end{document}